\documentclass[preprint,preprintnumbers,showpacs,amsmath,amssymb]{revtex4}
\usepackage{epsfig}
\begin{document}

\title{Magnetization damping in polycrystalline Co ultra-thin films:
Evidence for non-local effects}
\author{J-M. L. Beaujour, J. H. Lee, A. D. Kent} \affiliation{Department of Physics, New York University, 4
Washington Place, New York, NY 10003, USA}
\author{K. Krycka and C-C. Kao} \affiliation{Brookhaven National Laboratory, Upton, New York 11973, USA}
\date{\today}

\begin{abstract}
The magnetic properties and magnetization dynamics of
polycrystalline ultra-thin Co layers were investigated using a
broadband ferromagnetic resonance (FMR) technique at room
temperature. A variable thickness (1 nm $\leq t \leq$ 10 nm) Co
layer is sandwiched between 10 nm thick Cu layers (10 nm Cu$|t$
Co$|$10 nm Cu), while materials in contact with the Cu outer
interfaces are varied to determine their influence on the
magnetization damping. The resonance field and the linewidth were
studied for in-plane magnetic fields in field swept experiments at
a fixed frequency, from 4 to 25 GHz. The Co layers have a lower
magnetization density than the bulk, and an interface contribution
to the magnetic anisotropy normal to the film plane. The Gilbert
damping, as determined from the frequency dependence of the
linewidth, increases with decreasing Co layer thickness for films
with outer Pt layers. This enhancement is not observed in
structures without Pt layers. The result can be understood in
terms of a non-local contribution to the damping due to spin
pumping from Co through the Cu layer and spin relaxation in Pt
layers. Pt layers just 1.5 nm thick are found to be sufficient to
enhance the damping and thus act as efficient ``spin-sinks.'' In
structures with Pt outer layers, this non-local contribution to
the damping becomes predominant when the Co layer is thinner than
4 nm.
\end{abstract}

\maketitle

\section{introduction}
The magnetization dynamics of ultra-thin magnetic layers ($<$10
 nm) is of great scientific and technological interest, as such
layers are widely used in spin-injection and transport studies as
well as in magnetic devices. There has been particular interest in
non-local effects whereby layers separated from a magnetic layer
influence its magnetization dynamics through non-magnetic (NM)
metallic contact layers, i.e., via conduction electrons. Such
effects were modeled and studied in the early 1970's by Silsbee
\emph{et al.} \cite{silsbee}. More recently, a scattering theory
approach has been employed to describe the enhancement of the
damping \cite{brataas1,brataas2}. There have also been experiments
\cite{mizukami,heinrich,foros} which indicate quantitative
agreement with this theory based on interface parameters that can
be determined from \emph{ab-initio} calculations \cite{xia} as
well as transport experiments \cite{bauer}.

The current interest in this mechanism is at least threefold.
First, it is a fundamental mechanism of damping that can provide important
information on interface and bulk spin diffusion. Second, the
effect is known to play an important role in current-induced
magnetization excitations in spin-transfer devices \cite{sun}. In
such devices, a few nanometer thick magnetic layer is
embedded between NM layers which separate it from a thick
ferromagnetic layer that sets the spin-polarization of the current
\cite{nanopillar}. The threshold current density for magnetic
excitations is proportional to the damping \cite{sunPRB}. In order to understand
the physics of spin transfer, it is therefore important to
investigate the effect of adjacent NM layers on the magnetic
relaxation of ultra-thin films. Finally, from a technological
point of view, this process provides a way to engineer the
damping, which is important for high speed magneto-electronic
devices.

Mizukami \textit{et al.} studied the Gilbert damping of sputtered
NM$|t$ Py$|$NM films as a function of the FM layer thickness
($2\leq t \leq10$ nm) and for different adjacent non-magnetic
metals NM=Cu, Pd and Pt using a X-band FMR technique
\cite{mizukami}. The damping was found to be consistent with the
spin pumping picture: increasing with decreasing Py thickness for
the films with NM=Pt and Pd only. Further, the magnetization
damping of Cu$|$Py$|L$ Cu$|$Pt structures as a function of Cu
layer thickness $L$ and with fixed Py thickness, showed evidence
for a non-local effect. However, the non-local damping has been
studied mainly in NM$|t$ FM$|$NM structures as a function of the
FM layer thickness and varying the material directly in contact
with the FM layer \cite{mizukami, foros}. There have also been no
experimental studies of polycrystalline Co layers, which are
widely used in spin-transfer devices.

In this paper, we report systematic studies of the thickness
dependence of the linewidth and Gilbert damping of
ultra-thin Co layers in $||y_1$ Pt$|$Cu$|t$ Co$|$Cu$|y_2$ Pt$||$
structures. The thickness of the Cu layers in contact with Co is
kept fixed at 10 nm, which is chosen to be less than the spin-diffusion
length in Cu. The Pt layers have no direct interface with
the FM layer. The structure is modified by removing one or both of
the Pt layers ($y_1$ or $y_2=0$). The observation of changes in the
Gilbert damping confirm the non-local nature of this damping
mechanism and the data allow for quantitative analysis of the
interface spin-mixing conductances in the scattering theory
approach.

The paper is organized as follows. In section II, the film
fabrication and the FMR setup are described. Section III explains
the method of analysis of the resonance field and linewidth. In
section IV, the resonance field and the effective demagnetization
field are studied as a function of Co layer thickness. The FMR
linewidth and the Gilbert damping data are presented. This is
followed by a discussion of the dependence of the Gilbert damping
on the Co layer thickness for films with and without Pt and a
quantitative analysis of the data.

\section{Experimental technique}
Four series of films were fabricated with the layer structure
$||y_1$ Pt$|$10 nm Cu $|t$ Co$|$10 nm Cu$|y_2$ Pt$||$ on GaAs
substrates. First, symmetric structures with a variable thickness
of Co between Pt and Cu layers were grown with $y_1=y_2=1.5$ nm
and $1 \leq t \leq 10$ nm. The second set of samples were
asymmetric, without the top Pt layer ($y_1=1.5$ nm and $y_2=0$),
and with Co of 1.5, 2 and 2.5 nm thickness. In addition, films
without Pt layers ($y_1=y_2=0$), with Co of 2 and 3 nm thickness
were fabricated. In the fourth and final set of films, the
thickness of the Pt layers was varied from 0 to 5 nm in a
symmetric way with $y_1=y_2=0$, 1.5, 3 and 5 nm with a fixed Co
layer thickness of $t=2$ nm. Note that the studies in which Pt
layer thicknesses were varied focused on thin Co layers because
the non-local effect on the damping were found to be
significant only in Co layers thinner than 4 nm.

Films were prepared by a combination of electron-beam (Co, Pt) and
thermal (Cu) evaporation in an UHV system at a base pressure of $5
\times 10^{-8}$ Torr on polished semi-insulating $4 \times 6$
mm$^2$ GaAs wafers. Note that the chamber is equipped of an
in-situ wedge growth mechanism that enables the fabrication of a
number of samples with different Co or Pt thickness in the same
deposition run, without breaking vacuum. The evaporation rate for
Co and Cu were 0.5 and 0.8 \AA /sec respectively.

\begin{figure}
\begin{center}\includegraphics[width=8cm]{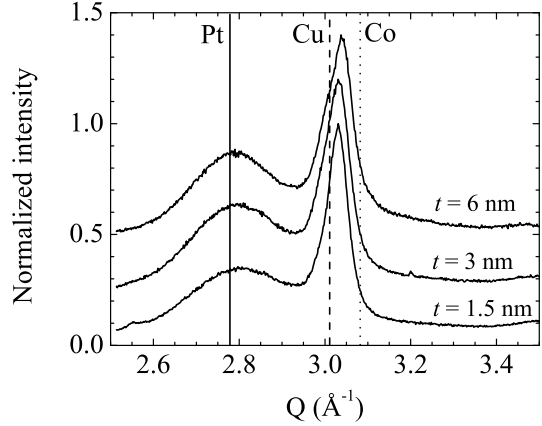}
\vspace{0 mm} \caption{X-ray data of $||$Pt$|$Cu$|t$
Co$|$Cu$|$Pt$||$ films with $t=$1.5, 3 and 6 nm. The curves for
the films with 3 and 6 nm Co layer were shifted up for clarity.
The vertical solid, dashed and dotted lines show the Q for nominal
fcc Pt, Cu and Co respectively.} \vspace{-5 mm}
\end{center}
\end{figure}

X-ray measurements were conducted on $||$Pt$|$Cu$|t$
Co$|$Cu$|$Pt$||$ with $t=1.5$, $3$ and $6$ nm (Fig. 1). The
reflectivity scans were taken at 19 keV using a Bicron point
detector. The intensity was first normalized by a beam-monitoring
ion chamber, and then each scan was rescaled to reach a maximum of
one. The films are polycrystalline and Pt, Cu and Co have an fcc
structure with (111) as the crystallographic growth direction.
Note that the peak at Q$\sim 3$ \AA$^{-1}$ shifts slightly between
the (111) nominal values for Co and Cu depending on how much Co is
present. An undistorted bcc (the most intense peak of 110 for Co
would be at 3.15 \AA$^{-1}$) is clearly not present, but a
distorted bcc with a lattice of 2.8 \AA \ cannot be ruled out.
Note that the peak of intensity at $Q\approx2.79$ \AA \ from the
Pt layer is strong evidence against interdiffusion at the Pt$|$Cu
interface.

The surface topography of the films $||$Pt$|$Cu$|$Co$|$Cu$||$ and
$||$Cu$|$Co$|$Cu$||$ were studied using non-contact Atomic Force
Microscopy (AFM) (Fig. 2). From the analysis of the AFM images, a
rms roughness of $\sigma=1.4\pm0.1$ nm was found for films with Pt
underlayer and $\sigma=1.3\pm0.2$ nm for films without Pt. The
lateral correlation length were $\xi=19.6\pm1.7$ nm and
$\xi=18.0\pm0.4$ nm respectively. There is no clear difference in
the rms roughness and the correlation length of the film with and
without Pt, suggesting that 1.5 nm Pt does not alter the growth of
the subsequent Cu$|$Co$|$Cu trilayers. The rms roughness is about
$6\%$ of the film thickness, which is typical of films prepared by
evaporation.

\begin{figure}
\begin{center}\includegraphics[width=8cm]{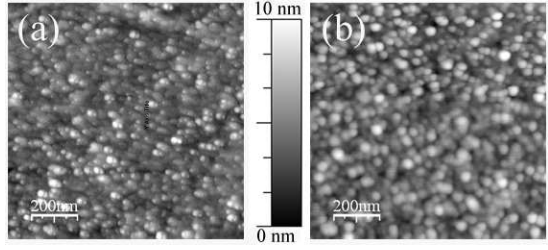}
\vspace{0 mm} \caption{Grey scale images of AFM scans performed on
1 $\mu m ^2$ areas of a magnetic film (a) $||$Cu$|$Co$|$Cu$||$ and
(b) $||$Pt$|$Cu$|$Co$|$Cu$||$ with a Co layer of 2 nm thickness.
The rms values of surface roughness is $\sigma=1.3\pm0.2$ nm for
(a) and $\sigma=1.4\pm0.1$ nm for (b), and the lateral correlation
length is $\xi=18.0\pm0.4$ nm and $\xi=19.6\pm1.7$ nm
respectively.} \vspace{-5 mm}
\end{center}
\end{figure}

FMR measurements were conducted in an in-plane field geometry at
room temperature employing a coplanar waveguide (CPW) as an ac
magnetic field generator and inductive sensor \cite{barry}. The
CPW was fabricated on a 350 $\mu$m thick semi-insulating and
polished GaAs wafer from a 200 nm thick Au film patterned using
bi-layer photolithographic process. It is characterized by a
transmission line of 50 $\mu$m width, a gap to the ground plate of
32 $\mu$m and a length of 4 mm, which is designed to have 50
$\Omega$ impedance above 4 GHz.  The CPW was placed into a brass
cavity, with its ends connected directly to the ports of a Network
Analyzer. Care was taken to avoid magnetic components in the
cavity and in all contacts to the CPW. FMR spectra were measured
by placing the magnetic sample metal face down on the CPW and
sweeping the external magnetic field at fixed microwave frequency
while measuring the S-parameters of the transmission line. Our
setup enables measurement of the FMR response of Co layers as thin
as 1 nm. Fig. 3a shows the geometry of the measurements. The
applied field produced by an electromagnet is directed along the
axis of the transmission line and perpendicular to the ac magnetic
field generated by the CPW. The applied field was in the film
plane, and was monitored with a Hall probe sensor that was
calibrated using electron paramagnetic resonance (EPR) on
2,2-diphenyl-1-picrylhydrazyl (dpph),  a spin 1/2 system. The
measured absorption of dpph is shown in Fig. 3b. The resonance
fields were always in agreement with the readings from the
Gaussmeter within 10-15 Gauss.

The FMR response was recorded at different frequencies in the
range 4-25 GHz. The spectra is measured as the relative change in
the transmitted power versus applied field. At 13 GHz for example,
the absorption from 5 nm thick Co film at resonance leads to a
0.66$\%$ decrease in the transmission. Thus, the susceptibility of
the magnetic films only causes a small change in the impedance of
the CPW and the absorption line can be analyzed as a small
perturbation to the CPW transmission.
\begin{figure}
\begin{center}\includegraphics[width=7cm]{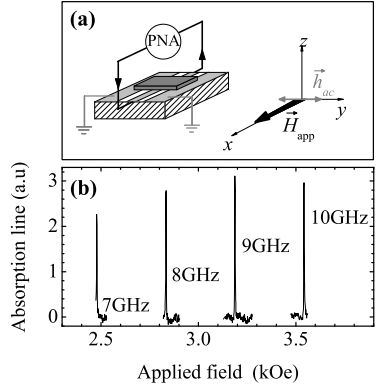}
\vspace{-2 mm} \caption{a) The FMR setup and field geometry. b)
Absorption of dpph at different frequencies. The resonance field
$H_{\text{res}}$ depends linearly on frequency: $H_{\text{res}}=(2
\pi \hbar / g \mu _B)f$, and is used to verify the calibration of
a Hall sensor used in this study.} \vspace{-5 mm}
\end{center}
\end{figure}

\section{method of analysis of the resonance field and linewidth}
The dynamics of the magnetization can be described in the
classical limit by the Landau-Lifshitz equation of motion:
\begin{equation}
\label{eq.1} {{1}\over{\gamma}}{{\partial \vec{M}}\over{\partial
t}}=\vec{M}\times \vec{H}_{\text{eff}} + {{\text{G}}\over{\gamma
^2 M_{\text{s}}^2}} \vec{M} \times {{\partial
\vec{M}}\over{\partial t}} \;,
\end{equation}
where $\vec{H}_{\text{eff}}$ is the effective field, $\vec{M}$ is
the magnetization vector and $\text{G}$ is the Gilbert damping
constant. The gyromagnetic ratio, $\gamma = g \mu _B / \hbar$, is
proportional to $g$, the Land\'{e} gyroscopic factor. For a film
magnetized in the film plane in an ac field, the resonance
condition is \cite{kittel}:
\begin{equation}
\label{eq.2} \left({{2 \pi f}\over{\gamma}} \right)^2 =
H_{\text{res}} \left(H_{\text{res}}+4 \pi M_{\text{eff}}
\right)\;,
\end{equation}
where for a continuous film, the effective demagnetization field
is given by:
\begin{equation}
\label{eq.3} 4 \pi M_{\text{eff}} = 4 \pi M_{\text{s}} + {{2
K_{\text{s}}}\over{M_{\text{s}}\ t}}\;.
\end{equation}
$M_{\text{s}}$ is the saturation magnetization. The uniaxial
anisotropy field $H_{\text{s}}=2K_{\text{s}} / M_{\text{s}}t$ is
characterized by a 1/t thickness dependence, where the anisotropy
originates from interface and/or strain-magnetoelastic
interactions. If $K_{\text{s}}$, the uniaxial anisotropy constant,
is negative, $H_\text{s}$ is directed out-of the film plane,
corresponding to a perpendicular component to the magnetic
anisotropy. Note that by assuming $M_{\text{s}}$ is independent of
thickness the uniaxial anisotropy constant, extracted from $4 \pi
M_{\text{eff}}$ versus $t$, can be overestimated.

The Gilbert damping is determined by the frequency dependence of
the FMR linewidth $\Delta H$ \cite{inhomogeneity}:
\begin{equation}
\label{eq.4} \Delta H(f) = \Delta H_0 + {{4 \pi
G}\over{\gamma ^2 M_{\text{s}}}}f\;,
\end{equation}
where the slope of $\Delta H(f)$ is the intrinsic contribution to
the linewidth, and is proportional to the Gilbert damping constant
$\text{G}$. $\Delta H_0$, the zero-frequency intercept, is usually
considered to be an extrinsic contribution to the linewidth.
$\Delta H_0$ is sensitive to the film quality: the highest quality
films typically exhibit a smallest residual or zero field
linewidth \cite{inhomogeneity, celinski}.

\section{Resonance field}
\begin{figure}
\begin{center}\includegraphics[width=7cm]{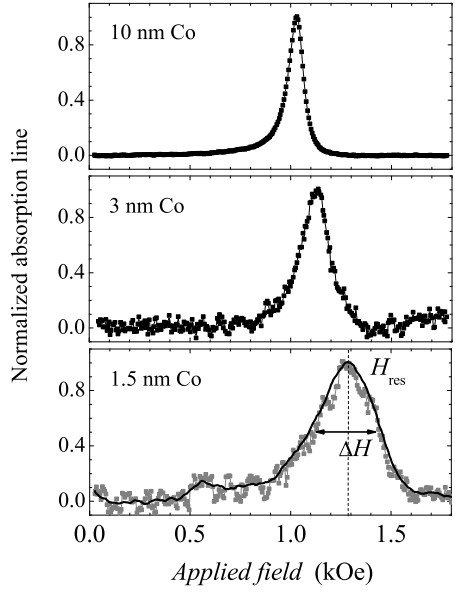}
\vspace{-2 mm}\caption{Typical absorption line at 13 GHz for
$||$Pt$|$Cu$|t$ Co$|$Cu$|$Pt$||$ film with $t=$10, 3 and 1.5 nm. With decreasing Co thickness
the absorption line shifts to higher field and broadens.}
\end{center}
\end{figure}
Fig. 4 presents the normalized FMR peak at 13 GHz for a selection
of $||$Pt$|$Cu$|t$ Co $|$Cu$|$Pt$||$ films. The absorption lines
were normalized by substracting the background signal and dividing
by the relative change in transmission at resonance. The lineshape
of the FMR curves is typically Lorentzian. Note that we observed
asymmetric lineshapes at some frequencies, associated with the
measurement method. Distortion of the FMR lineshapes originates
from the mixing of the absorptive and dispersive components of the
susceptibility (see Ref. \cite{kalarickal}). Fig. 5 shows the
thickness dependence of the resonance field $H_{\text{res}}$ at 14
GHz for films with and without 1.5 nm Pt. The resonance field is
practically constant ($H_{\text{res}} \approx 1.2$ kOe) when the
Co layer thickness is larger than 5 nm. For thinner layers,
$H_{\text{res}}$ increases with decreasing $t$, reaching a value
of 1.6 kOe for the film with 1 nm Co layer. The films with the
same thickness of Co exhibit about the same resonance field,
independent of the presence of the Pt layers. The run-to-run
variations in magnetic properties from small changes in film
deposition conditions, for instance, are significantly less than
the changes in $H_{\text{res}}$ observed in the very thin film
limit.

\begin{figure}
\begin{center}\includegraphics[width=7cm]{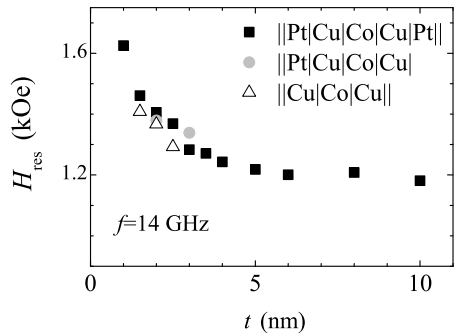}
\vspace{-2 mm}\caption{Thickness dependence of the resonance field
at 14 GHz.}
\end{center}
\end{figure}

The resonance field was measured at different frequencies (Fig.
6). The effective demagnetization, $4 \pi M_{\text{eff}}$, and the
g-factor were found by fitting $f^2/H_{\text{res}}$ $vs.$
$H_{\text{res}}$ to Eq. 2. The slope of $f^2/H_{\text{res}}$ gives
the g-factor and the zero frequency intercept provides the
effective demagnetization field (see the inset of Fig. 6). Films
with equal Co layer thickness have nearly the same $g$ and $4 \pi
M_{\text{eff}}$. This suggests that the Pt underlayer and capping
layer do not affect the magnetic properties of the films.
\begin{figure}
\begin{center}\includegraphics[width=7cm]{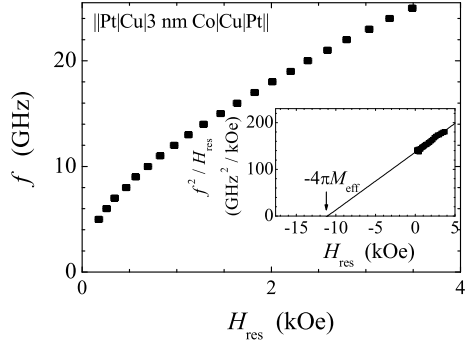}
\vspace{-2 mm}\caption{The frequency dependence of the resonance
field for a $||$Pt$|$Cu$|t$ 3 nm Co $|$Cu$|$Pt$||$ film. The inset
shows $f^2/H_{\text{res}}$ $vs.$ $H_{\text{res}}$.}
\end{center}
\end{figure}
The effective field exhibits a clear thickness dependence,
decreasing from about 15 kOe to about 8 kOe when the Co thickness
varies from 10 nm to 1 nm (Fig. 7). The thickness dependence of
the effective field can be understood as originating from the
presence of an uniaxial anisotropy field that depends on thickness
as 1/$t$. The magnetization density is assumed to be independent
of the film thickness, as Co and Cu are immiscible. The best fit
to Eq. 3 gives a saturation magnetization density of
$M_{\text{s}}=1131$ emu/cm$^3$ and an uniaxial anisotropy constant
$K_{\text{s}}=-0.46$ erg/cm$^2$. The value of $M_{\text{s}}$ is
smaller than the magnetization density of bulk fcc Co
($M_{\text{s}}$=1400 emu/cm$^3$). The negative sign of
$K_{\text{s}}$ reflects a perpendicular component of magnetic
anisotropy. As noted earlier, the origin of this perpendicular
anisotropy may be a magnetic anisotropy associated with Co$|$Cu
interfaces or magnetoelastic interactions associated with strain
in the Co layers, which increases with decreasing Co thickness.
The results are in good agreement with magnetometry studies of 150
nm thick epitaxial Co$|$Cu multilayers grown on GaAs substrates,
with Co layer thickness ranging from 0.5 to 4 nm \cite{ohandley}.
The authors found an average magnetization density
$M_{\text{s}}$=1241 emu/cm$^3$ and an anisotropy constant
$K_{\text{s}}=-0.47$ erg/cm$^2$. The weighted average value of the
Land\'{e} g-factor is $g=$2.49$\pm$0.14 (inset of Fig. 7). This is
larger than the value reported in the literature for fcc phase Co
($g$=2.15) \cite{wiedwald}. Note that in thin layers there is a
larger uncertainty in $g$ making it difficult to determine the
thickness dependence of the g-factor.

\begin{figure}
\begin{center}\includegraphics[width=7cm]{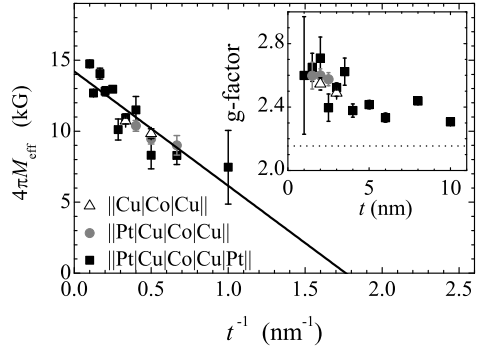}
\vspace{-2 mm}\caption{The effective demagnetization field $4 \pi
M_{\text{eff}}$ $vs.$ 1/$t$. The solid line is the best fit based
on Eq. 3. The inset shows the thickness dependence of the
Land\'{e} $g$ factor. The dashed line gives the value of $g$ for
fcc bulk (2.15).}
\end{center}
\end{figure}

\section{Linewidth and Gilbert damping}
The linewidth was studied as a function of frequency and Co layer
thickness. As mentioned earlier (section IV), we observed
asymmetry in the FMR lineshape at some frequencies. In those
cases, the half power linewidth was extracted using the procedure
described in Ref. \cite{kalarickal}. The linewidth at 10 and 14
GHz is plotted versus the thickness in Fig. 8 for the films with
and without Pt. For thick films ($t\geq 5$ nm) with two Pt outer
layers, $\Delta H$ is practically independent of thickness.
However, the linewidth of thinner films ($t < 5$ nm) increases
strongly with decreasing Co layer thickness. The three series of
films show different increases of the linewidth for thin layers.
For instance, the film with 2 nm Co layer thickness and no Pt has
a linewidth of 120 Oe at 10 GHz. This is smaller than that of the
film with a Pt underlayer ($\Delta H$=180 Oe), and that with two
Pt layers ($\Delta H$=260 Oe).
\begin{figure}
\begin{center}\includegraphics[width=7cm]{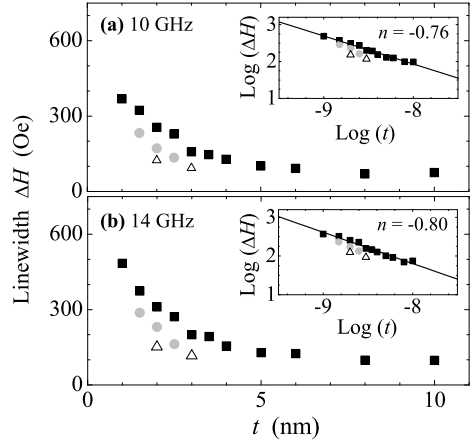}
\caption{Thickness dependence of the linewidth at 10 and 14 GHz
for the films of series without and with Pt. The insets show the
corresponding data on a log-log scale.}
\end{center}
\end{figure}
\begin{figure}
\begin{center}\includegraphics[width=7cm]{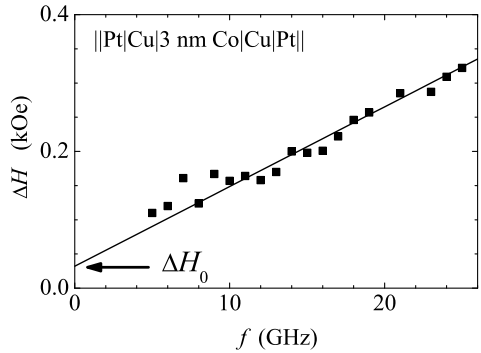}
\caption{Typical frequency dependence of the linewidth for the
film $||$Pt$|$Cu$|$Co$|$Cu$|$Pt$||$ with Co layer thickness of 3
nm. $\Delta H_0$ and $d \Delta H / df$ are extracted from
the linear best fit (solid line).}
\end{center}
\end{figure}
The frequency dependence of $\Delta H$ enables a determination of
the Gilbert damping and the inhomogeneous contribution to the
linewidth. Fig. 9 shows the frequency dependence of $\Delta H$ for
a symmetric film with Pt layers ($y_1=y_2=1.5$ nm) and a 3 nm
thick layer of Co. The linewidth depends linearly on frequency,
with a zero frequency offset. Below 10 GHz, the data points are
more scattered. In this frequency range, the resonance field is of
the order of or smaller than the film saturation field, and the
absorption line becomes asymmetric. A linear fit to the data is
shown in Fig. 9. The thickness dependence of the slope $d \Delta H
/ df$ and intercept $\Delta H_0$ are shown in Fig. 10. The two
parameters exhibit similar thickness dependence: decreasing with
increasing Co layer thickness. Therefore, both changes in the
Gilbert damping and inhomogeneous broadening contribute to the
enhanced linewidth for thin layers. $\Delta H_0$ approaches zero
for thick Co reflecting the good quality of these layers. The
slope is also practically constant when $t \geq 5$ nm, and it is
about three times larger for the thinnest films with two Pt
layers.
\begin{figure}
\begin{center}\includegraphics[width=7cm]{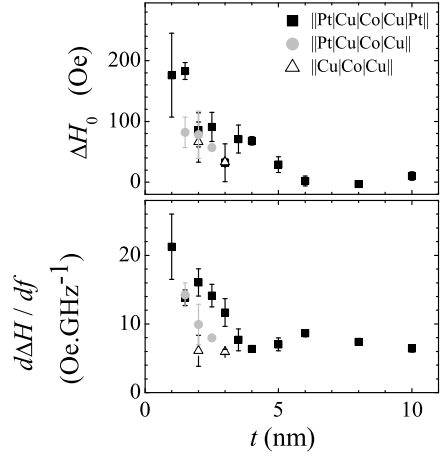}
\caption{Thickness dependence of the two contributions to the
linewidth: $\Delta H_0$ and $d \Delta H / df$}
\end{center}
\end{figure}

The Gilbert damping G was estimated from $d \Delta H / df$ (Fig.
10) with Eq. 4. The thickness dependence of the damping is shown
in Fig. 11 for the films with and without Pt layers. The damping
of $||$Cu$|$Co$|$Cu$||$ films is practically independent of the
thickness, and it is about equal to the damping of the films with
two Pt layers and thick Co layers ($t>5$ nm). Note, however, that
the inhomogeneous contribution to the linewidth increases with
decreasing Co thickness (Fig. 10). The films with one or two Pt
layers show an increase in the damping when the Co layer is
thinner than 4 nm. 1.5 nm thick Pt layers are thus sufficient to
lead to the enhancement of the damping.

In order to investigate whether the damping is a function of the
Pt outer layers, films with 2 nm thick Co layers and variable Pt
layer thickness were studied. The film structure was $||y$
Pt$|$Cu$|$2 nm Co$|$Cu$|y$ Pt$||$. The results are shown in the
inset of Fig. 11. The films without Pt have two times smaller
damping than the films with Pt. However, remarkably, the damping
does not increase further as the Pt layer thickness is increased
beyond 1.5 nm. It saturates at a value of about $6 \times 10^8$
s$^{-1}$. This result clearly shows that the main origin of the
enhancement of the damping in thin Co layers are the Cu$|$Pt
interfaces. Note that the films of this series, including the ones
with $y=0$ and $y=1.5$ nm were fabricated a few months after the
other series of films. It can be seen that the Gilbert damping of
the samples with $y=0$ and $y=1.5$ nm, plotted in the inset, are
consistent with the films deposited earlier (shown in the main
part of Fig. 11).
\begin{figure}
\begin{center}\includegraphics[width=7cm]{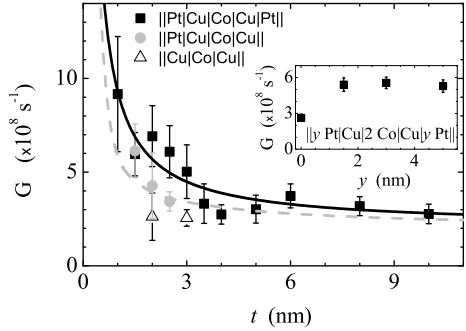}
\caption{The magnetic damping versus ferromagnetic layer
thickness. The solid line is the best fit of Eq. 5 to G($t$) for
the films with two Pt layers. The dashed grey line is the
calculated thickness dependence of G for the films with a single
Pt$|$Cu interface, and using the parameters found in the earlier
fitting. Inset: the magnetic damping versus the thickness of the
Pt layers for the films $||y$ Pt$|$Cu$|$Co$|$Cu$|y$ Pt$||$ with 2
nm Co layer.}
\end{center}
\end{figure}

\section{Discussion and interpretation}
There are a number of mechanisms which lead to the broadening of FMR linewidth
in very thin magnetic films. In the present study,  particularly from the effect of
``remote'' Pt layers on the linewidth, it is clear that non-local interactions are important. We
discuss this in detail below and follow this with a brief discussion of two-magnon
scattering, which has been extensively discussed in the context of damping in ultra-thin films.

The dependence of $\text{G}$ on the thickness of the FM and on the
presence of the Pt can be understood in term of spin pumping
induced enhancement of the magnetic damping. In this model, the
precessing magnetization of the thin Co layer generates a spin
current that flows through the Cu layers. In absence of Pt
($||$Cu$|$Co$|$Cu$||$ system), there is not a significant
additional damping with decreasing Co layer thickness because the
Cu layers are 10 nm thick, which is much thinner than the spin
diffusion length in Cu ($\lambda _{\text{sd}} \approx$ 350 nm at
room temperature \cite{jedema}). When the Cu layer is adjacent to
Pt, a strong spin scatterer, the spin current is absorbed in Pt.
There is no back flow of spin current. As a consequence, the
Gilbert damping of the FM is enhanced. For a symmetric film with
two Pt layers, the effect is more pronounced because the pumping
current is absorbed at the two interfaces with Pt. Furthermore,
above 1.5 nm Pt, the Gilbert damping does not depend on the
thickness of Pt. In assuming that the film is continuous, a Pt
layer as thin as 1.5 nm is thus sufficient to act as a perfect
spin sink, i.e., the entire spin current is absorbed. This is in
agreement with the results in Ref. \cite{bass}, where it was
reported that the spin loss parameter at the Pt$|$Cu interface is
0.9 $\pm 0.1$, which implies about 70$\%$ spin randomization at
the interface. For very thin Co layers ($t\leq 4$ nm), the
non-local damping becomes predominant. It must be mentioned that
for the films without Pt capping layer, the top Cu layer, which is
in contact with air, might be covered by a native oxide layer. The
FMR measurements were conducted within a few days of fabrication
of the films. While $a$ $priori$ the native oxide layer may affect
the damping, our results indicate that this oxide layer must not
play a very important role. Indeed, assuming that the oxide layer
acts as a spin scatterer, like Pt, it would then be expected that
the damping $d \Delta H/df$ of $||$Pt$|$Cu$|$Co$|$Cu$||$ and
$||$Pt$|$Cu$|$Co$|$Cu$|$Pt$||$ with the same Co layer thickness
will be equal. Fig. 10 shows that this is not the case.

In a symmetric structure $||$NM$_2|$NM$_1|$FM$|$NM$_1|$NM$_2||$,
where NM$_2$ is a perfect spin sink and NM$_1$ thickness is much
smaller than the spin diffusion length of the material, the
Gilbert damping is \cite{brataas2}:
\begin{equation}
\label{eq.5} \text{G}(t) = \text{G}_0 + \left({{g \mu
_B}\over{e}}\right)^2 {{\tilde {\text{G}}_{\text{eff}}^{\uparrow
\downarrow} S^{-1}}\over{t}}\;.
\end{equation}
$\text{G}_0$ is the residual Gilbert damping (bulk damping), $S$
is the surface area of the sample and $\tilde
{\text{G}}_{\text{eff}} ^{\uparrow \downarrow}$ is the effective
spin mixing conductance. $\tilde{\text{G}}_{\text{eff}} ^{\uparrow
\downarrow}$ accounts for the spin mixing conductance at the
Co$|$Cu and Cu$|$Pt interfaces:
\begin{equation}
\label{eq.6} {{1}\over{\tilde{\text{G}}_{\text{eff}}^{\uparrow
\downarrow}S^{-1}}} =
{{1}\over{\tilde{\text{G}}_{\text{Co}|\text{Cu}}^{\uparrow
\downarrow}S^{-1}}} + {{1}\over{{\text{G}}_{\text{Cu}}}} +
{{1}\over{\tilde{\text{G}}_{\text{Cu}|\text{Pt}}S^{-1}}} \;,
\end{equation}
where $\text{G}_{\text{Cu}}=1/(2L \rho)$ is the conductance per
spin of the Cu layer of thickness $L$ and resistivity $\rho$.
Using the literature value of the resistivity for pure Cu, $\rho =
1.7 \times 10^{-8}\  \Omega$m, the conductance per spin of 10 nm
Cu layer is $\text{G}_{\text{Cu}}= 2.94 \times 10^{15}$ $\Omega
^{-1}$m$^{-2}$. In the films $||$Pt$|$Cu$|$Co$|$Cu$|$Pt$||$
studied here, the Cu thickness (10 nm) is much smaller than the Cu
spin diffusion length. In addition, we found that a 1.5 nm Pt
layer is sufficient to saturate the additional Gilbert damping,
and therefore is a perfect spin sink. The best fit of the
thickness dependence of the magnetic damping gives
$\tilde{\text{G}}_{\text{eff}} ^{\uparrow \downarrow}S^{-1} =
(0.34 \pm 0.05) \times 10^{15}$ $\Omega ^{-1}$m$^{-2}$ and
$\text{G}_0=2.09 \pm 0.44$ s$^{-1}$, using the average value of
the g-factor reported in section IV. Note that the conductance per
spin of a 10 nm Cu layer is about 10 times larger than the
estimated effective spin mixing resistance, meaning that the main
contributions to the resistance in the layered structures
originates from that associated with the spin mixing at the
Cu$|$Co and at the Cu$|$Pt interfaces. After correction of the
Sharvin conductance \cite{brataas2}, the effective spin mixing
conductance is $\text{G}_{\text{eff}} ^{\uparrow
\downarrow}S^{-1}=0.26 \pm 0.04 \times 10^{15}$ $\Omega ^{-1}
\text{m}^{-2}$. Using the theoretical values of the mixing
conductance for Cu$|$Co ($0.55 \times 10^{15}$ $\Omega
^{-1}$m${^{-2}}$) and Cu$|$Pt ($1.36 \times 10^{15}$ $\Omega
^{-1}$m$^{-2}$) given in \cite{brataas2} and the conductance per
spin of Cu as calculated above, the effective spin mixing
conductance of the films is predicted to be $0.34 \times 10^{15}$
$\Omega ^{-1}$m$^{-2}$. The experimental value of the effective
spin mixing is about 30$\%$ smaller than the calculated value. A
similar FMR study was conducted on sputtered films with the same
structure. An effective spin mixing conductance
$\text{G}_{\text{eff}} ^{\uparrow \downarrow}S^{-1}=0.41 \pm 0.05
\times 10^{15}$ $\Omega ^{-1} \text{m}^{-2}$ was found. The result
implies that the interface spin mixing conductance depends on the
film deposition method.

We now briefly discuss FMR line broadening due to two magnon
scattering. The scattering of magnons by defects and imperfections
at the surface and interface of a thin ferromagnetic film produces
modes that are degenerate with the FMR mode ($k$=0)
\cite{twomagnon}, leading to an additional contribution to the
linewidth. The mechanism is operative when the magnetization
vector lies in the film plane, and it is not operative in the
perpendicular geometry. Recently, Arias and Mills developed a
theory of the two-magnon scattering contribution to the FMR
linewidth in ultra-thin films \cite{mills}. The theory was found
to successfully explain the frequency dependence of the linewidth
of epitaxially grown thin films, with defects of rectangular shape
\cite{lindner, urban}. The AFM images (Fig. 2) show that there is
no evidence of anisotropy in the topology of the defects at the
surface of the polycrystalline films. In this case, according to
\cite{mills}, the two-magnon contribution is proportional to
$H_{\text{res}}^2$ and $H_{\text{s}}^2$. As shown in Fig. 9, the
linewidth increases linearly with the frequency in the range 4-25
GHz. In plotting the data as $\Delta H$ versus $H_{\text{res}}$,
the trend is similar to a square root dependence, which does not
agree with the expected $H_{\text{res}}^2$ dependence predicted by
\cite{mills}. Furthermore, the authors calculated that one of the
signatures of the linewidth broadening from two-magnon scattering
is $\Delta H \propto H_{\text{s}}^2$, where the coefficient of
proportionality contains information related to the roughness of
the film. As the uniaxial anisotropy $H_{\text{s}}$ scales as
$1/t$ (Eq. 3), it is predicted that the linewidth increases as
$1/t^2$ for the very thin films. The insets of Fig. 8a and 8b
shows the linewidth versus the Co layer thickness in log-log plot
at two frequencies, $f=10$ and $f=14$ GHz respectively. From the
linear best fit, we found that $\Delta H$ increases as $1/t^{n}$
with $n \approx 0.8$. We can conclude that the Arias and Mills
model of two-magnon scattering contribution to the linewidth
cannot be used to explain the thickness dependence of $\Delta H$.
While, in general a two-magnon contribution to the linewidth may
be present, it cannot explain the strong increase of the linewidth
for very thin films. This can be seen by comparing the films with
two Pt layers and those with only the Pt underlayer. The films
grown with a Pt underlayer ($y_1=1.5$ nm) and with or without top
Pt ($y_2=1.5, 0$ nm) are expected to have similar microstructure,
since the underlayer structure is identical ($||$1.5 nm Pt$|$10 nm
Cu). As a consequence, in the picture of two-magnon scattering
induces linewidth broadening, the linewidth of those films should
be the same. However, the thickness dependence of $\Delta H$ shows
that the films with two Pt layers have a larger linewidth than
those with a single Pt layer.

\section{conclusion}
We have conducted a FMR study of polycrystalline ultrathin Co
films embedded between 10 nm Cu. The outer interfaces of Cu were
placed in contact with different environments, by adding Pt
layers. We have found that very thin Co layers magnetization
relaxation processes depend on the non-local environment. The Co
layers exhibit a lower magnetization density compared to the bulk
material, and an uniaxial anisotropy field perpendicular to the
film plane. The large difference between the g-factor value of the
ultra-thin Co layer compared to the bulk fcc Co remains an open
question.

The FMR linewidth, studied as a function of the Co layer thickness
and of the non-local environment, increases with decreasing Co
layer thickness. We have provided evidence that the two-magnon
scattering mechanism cannot explain the thickness dependence of
the linewidth. The inhomogeneous contribution to $\triangle H$
increases with decreasing Co layer thickness, independently of the
presence of the Pt layer. However, the Gilbert damping, the
intrinsic contribution to the linewidth, was found to be function
of the Co layer thickness and depends on the non-local
environment. In particular, the Gilbert damping increases
significantly with decreasing Co layer thickness only in the
presence of a Pt$|$Cu interface. The thickness dependence of G is
consistent with the theory of spin pumping. By changing the outer
interface environment of $|$Cu$|$Co$|$Cu$|$ by adding or removing
Pt in contact with Cu, we gave unambiguous experimental evidence
that the damping enhancement is a non-local mechanism. The
non-local damping becomes predominant compared with other
relaxation mechanism when the Co layer is a few nanometer thick
($t\leq$4 nm).

In summary, the adjacent layers material as well as the non-local
environment are critical for understanding the magnetization
relaxation in ultra-thin FMs. This important result must be taking
into account in the study of spin transfer devices.

We appreciate very stimulating discussions with C. E. Patton and
J. Z. Sun. This research is supported by NSF-DMR-0405620 and by
ONR N0014-02-1-0995.

\end{document}